\documentclass[a4paper,11pt]{article}
\usepackage{graphicx}
\usepackage{subfigure}
\usepackage{graphicx,epsfig}
\usepackage{epstopdf}
\usepackage{latexsym}
\usepackage{amsfonts,amsmath,amssymb}
\usepackage{hyperref}
\usepackage{float}

\usepackage{epsfig,multicol,bbm}

\newcommand\fverb{\setbox\pippobox=\hbox\bgroup\verb}
\newcommand\fverbit{\egroup\item[\fbox{\unhbox\pippobox}]}

\newbox\pippobox

\begin{document}
\title{\bf Boosted   Cylindrical  Magnetized Kaluza-Klein Wormhole}
\author{S. Sedigheh Hashemi   \,\,and \,\,  Nematollah Riazi\thanks{Electronic address: n\_riazi@sbu.ac.ir} 
\\
\small Department of Physics, Shahid Beheshti University, G.C., Evin, Tehran 19839,  Iran}
\maketitle
\begin{abstract}
In this work, we consider  a vacuum solution of Kaluza-Klein theory with cylindrical symmetry.  We
investigate the physical properties of the solution as viewed in 
  four dimensional spacetime, which turns out to be a
 stationary, cylindrical wormhole supported by a scalar field and a magnetic field oriented along the  wormhole.  
  We then apply a boost to the five dimensional solution along the extra dimension, and perform the Kaluza-Klein reduction. As a result, we show that the new solution is  still a  wormhole with a radial electric field and a magnetic field stretched along the wormhole throat.
\end{abstract}
\section{Introduction}
Wormholes are  topological structures like bridges or tunnels that connect different universes or different parts of the same universe. They were first suggested by Einstein and Rosen \cite{ER}, who noted that the Schwarzschild black hole has two exterior asymptotic regions connected by a throat (or bridge).
 The Einstein-Rosen wormhole is spacelike and  classical objects can not pass through it, however, it has been argued that it  can perhaps connect quantum particles in order to produce quantum entanglement and also the Einstein-Pololsky-Rosen effect \cite{EPR}, \cite{1306.0533}.
Wormholes might thus provide a  geometric description of elementary particles,  perhaps, at the Planck scale \cite{MW}. On the other hand, wormholes can be used to describe initial data for the Einstein equations \cite{Misner}, \cite{1411.1084} whose time evolution is equivalent to the black  hole collisions that has been recently observed by LIGO \cite{1602.03837}. Further, they can supply information  for understanding the evaporation of black  holes \cite{SW},
  and also the  evolution  of universe \cite{SW}-\cite{DH}. Additionally, the microscopic wormholes could present a mechanism for  vanishing cosmological constant problem \cite{SC}-\cite{IK}.

An interesting topic is the general theory of a traversable Lorentzian wormholes which were introduced by Morris and Thorne \cite{MT}. These wormholes are static and spherically symmetric bridges accessible for classical particles or light \cite{V}. Moreover, the static wormholes can be the solutions of Einstein equations in the presence of exotic matter such as phantom fields with negative kinetic energy  which violate the null energy condition \cite{KB, Ellis}. 
Henceforth, for solving this problem, wormholes can be studied in other alternative theories of gravity such as
 the Gauss-Bonnet theory \cite{1108.3003}, the theories of non-minimally coupled scalar fields \cite{1111.3415},  massive (bi)gravity \cite{1502.03712},
and  higher dimensional  theories of gravity \cite{0212112}. Among possible extensions of higher dimensional models, in this paper, we will focus on the Kaluza-Klein theory, in which  the existence of an extra dimension would give rise to a unified picture of general relativity and electromagnetism \cite{kaluza}.

In \cite{SY}, the authors obtained an analytical solution of wormholes in Kaluza-Klein theory with the topology of spacetime being $R^1\otimes S^3 \otimes M^d$, in which the dimension of spacetime equals $D=1+3+d$, with $d$ being the dimension of internal space. They supposed that the internal space is static and compact, with a time dependent four dimensional spacetime. 
Chodos and Detweiler \cite{CD}, obtained a class of spherically symmetric and asymptotically flat solutions which described wormholes in Kaluza-Klein theory. Furthermore, their solution was expanded to axisymmetric multi-wormholes \cite{GC}. 
In \cite{SI}, and  \cite{SID}, it can be seen that, by parametrizing the off-diagonal metric elements in five dimensional Kaluza-Klein theory, the new locally anisotropic wormholes, which are vacuum solutions of Einstein field equations will be defined. In addition, cylindrically symmetric Abelian wormholes in $(4+n)$ dimensions in the context  of Kaluza-Klein theory are obtained in \cite{AV}.

In this paper, we consider the magnetized and cylindrical Kaluza-Klein wormhole, which is a vacuum solution of Einstein field equations in five dimensions \cite{ssh}. We apply a boost along the extra dimension, and by performing the Kaluza-Klein reduction we arrive at a four dimensional solution, which is still a wormhole. However, the boosted solution has both an  electric field normal to the wormhole and a  magnetic field stretched along the wormhole throat. Consequently, we investigate the behavior of the magnetic and electric flux parallel and through the wormhole. 
 We then explore the gauge fields, which are  due to the topology of the spatial compact manifold in a higher dimensional manifold.

The organization of the paper is as follows. We begin in section  \ref{r} by reviewing and briefly discussing  the cylindrical and magnetized wormhole obtained in \cite{ssh}. We next present in section \ref{o} the boosted Kaluza-Klein magnetized wormhole, and perfom a Kaluza-Klein reduction to obtain  a new four dimensional wormhole, and investigate its properties. In section \ref{M} we will calculate the magnetic and electric flux through the wormhole.
 Section \ref{s} concludes with some comments.

\section{The  Kaluza-Klein Wormhole: a brief review}\label{r}
We begin by reviewing the original    magnetized cylindrical wormhole which is a vacuum solution of Einstein field equations in five dimensions in the context of Kaluza-Klein theory (see \cite{ssh} for derivations). The solution is  
\begin{equation}\label{eq1}
{\rm d}s^2_{(5)}=-\frac{c}{(ar)^{2/3}}{\rm d}t^2+{\rm d}r^2+{\rm d}z^2-r^{4/3}|\left(c+d\ln r\right)|{\rm d}w^2+2(ar)^{4/3}{\rm d}w{\rm d}\theta,
\end{equation}
where $r$ is cylindrical radial coordinate, $z  \in   (-\infty, +\infty)$ is the longitudinal coordinate,  $\theta \in [0, 2\pi]$, and $w$ is the extra coordinate. $a$, $c$, and $d$ are parameters.

By performing a Kaluza-Klein reduction \cite{sorkin, gross} along the coordinate $w$, the following scalar, and  gauge fields can be obtained
 \begin{equation}
  \phi^2=|r^\frac{4}{3}\left(c+d\ln r \right)|,\quad A_{\theta}=\frac{a^\frac{4}{3}}{\kappa |c+d\ln {r}|}.
  \end{equation}
 One can  immediately find the electromagnetic tensor field as
  \begin{equation}
  F_{r \theta}=-F_{\theta r}=-\frac{da^\frac{4}{3}}{\kappa r\left(c+d\ln r\right)^2},
  \end{equation}
which corresponds to a magnetic field along the $z$ coordinate $(B_z)$. Furthermore, the four-dimensional spacetime can be calculated via the reduction from the five dimensional metric (\ref{eq1}), given by
\begin{equation}\label{eq12}
{\rm d}s^2_{(4)}=-\frac{c}{(ar)^\frac{2}{3}  }{\rm d}t^2+ {\rm d}r^2+
\frac{a^\frac{8}{3} r^\frac{4}{3}}{|c+d\ln{r}|}{\rm d}\theta^2+{\rm d}z^2,
\end{equation}
which is  a static, cylindrical wormhole supported by a scalar field and a magnetic field oriented along the  wormhole ($z$-axis).  

To conclude this section, we may remark that  the magnetic flux on either side of the throat  converge to a finite value. For understanding the gravitational and electromagnetic effects of the wormhole,   the equations of motion for a neutral and a charged test particle were calculated, and we found the  repulsive character of the wormhole gravitational field. Finally,  we showed  that the null energy condition  is violated for this solution \cite{ssh}.

\section{The Boosted Kaluza-Klein Wormhole}\label{o}
In this section, we perform a boost to the Kaluza-Klein magnetized cylindrical wormhole (\ref{eq1}). The applied boost is along the extra coordinate $w$ with the boost parameter $\alpha$.
We can make the following coordinate boost by renaming the metric coordinates  (\ref{eq1}) as $(t',r', z', \theta', w' )$ and representing the boosted coordinate as $(t, r, z, \theta, w)$
\begin{eqnarray}
t^{\prime}&= &t\cosh \alpha -w \sinh \alpha~, \\
w ^{\prime}&= &w\cosh \alpha-t \sinh \alpha,
\end{eqnarray}
with these transformations, the metric (\ref{eq1})  will become
\begin{align}\label{7}
{\rm d}s^2=&-\left(\frac{c}{(ar)^{2/3}}   \cosh^2\alpha+   r^{4/3}|\left(c+d\ln r\right)|\sinh^2 \alpha       \right){\rm d}t^2
+{\rm d}r^2+{\rm d}z^2\nonumber\\&
-\left(\frac{c}{(ar)^{2/3}}   \sinh^2\alpha+   r^{4/3}|\left(c+d\ln r\right)|\cosh^2 \alpha       \right){\rm d}w^2
\nonumber\\& +\left(                 \frac{c}{(ar)^{2/3}}+  r^{4/3}|\left(c+d\ln r\right)|  \right)\sinh 2\alpha  {\rm d}t{\rm d}w
\nonumber\\&-2(ar)^{4/3}\sinh \alpha {\rm d}\theta{\rm d}t
\nonumber\\&+2(ar)^{4/3}\cosh \alpha {\rm d}\theta{\rm d}w,
\end{align}
which again is  a stationary (but not static) vacuum solution of Einstein field equations with  vanishing Ricci scalar and Ricci tensor.

For our new metric (\ref{7}), we can perform the Kaluza-Klein reduction on the extra coordinate $w$, and using  the following ansatz  for the metric which lies in the premise that the compact dimension of a $N$ dimensional differential  manifold is orthogonal  to  the manifold
\cite{sorkin}-\cite{1609.09045}
\begin{equation}\label{eq8}
(\hat g_{AB})= \left(
‎\begin{array}{ccccccc}‎
‎g_{\mu \nu}+\kappa^2 \psi A_{\mu} A_{\nu}‎   & \kappa\psi A_{\mu} &  \\‎ 
 ‎‎  &‎\\‎ 
\kappa \psi A_{\nu}    ~~&\psi&  \\‎ 
‎\end{array}
‎\right),
\end{equation}
where $\hat g_{AB}$ is the five dimensional metric,  and the four dimensional spacetime is given by $g_{\mu \nu}$. $A_{\mu}$, and $\psi$ are the vector (gauge) and scalar fields as new gravitational degrees of freedom, respectively.  Therefore, by using (\ref{eq8}) we  will obtain the four dimensional  metric with the following components
\begin{align}
g_{tt}&=-\frac{c \cosh ^2\alpha }{(a r)^{2/3}}-r^{4/3} \sinh ^2\alpha|(c+d \ln r)|\nonumber\\&
+
\frac{\sinh ^22 \alpha  \left(r^{4/3} (a r)^{2/3} |(c+d \ln r)|+c\right)^2}{4 (a r)^{2/3}
   \left(r^{4/3} (a r)^{2/3} \cosh ^2\alpha  |(c+d \ln r)|+c \sinh ^2\alpha
 \right)},
\end{align}
\begin{equation}
g_{rr}=g_{zz}=1,
\end{equation}
\begin{equation}
g_{\theta \theta}=\frac{a^4 r^4 \cosh ^2\alpha }{(a r)^{2/3} \left(r^{4/3} (a r)^{2/3} \cosh ^2\alpha |(c+d
   \ln r)|+c \sinh ^2\alpha \right)},
\end{equation}
and
\begin{align}
  g_{t\theta}=\frac{a^2 r^2 \sinh 2 \alpha \cosh \alpha  \left(r^{4/3} (a r)^{2/3} |(c+d \ln r)|+c\right)}{2 (a r)^{2/3} \left(r^{4/3} (a r)^{2/3} \cosh ^2\alpha |(c+d \ln r)|+c
   \sinh ^2\alpha \right)}-(a r)^{4/3} \sinh \alpha.
\end{align}
It is well known that a boost along the extra dimension, when applied to a  five dimensional Kaluza Klein solution, leads to electrically charged solutions in the projected $(3+1) $  dimensional solution. For example, if we start with a neutral, Schwarzschild-like solution in five dimensions, the boosted solution becomes something like the Reissner-Nordstr\"{o}m charged black hole \cite{1107.5563}. 

The boost along the fifth dimension, in the wormhole case we are considering, has led to a charged and magnetic solution. It is seen that a cross term $ {\rm d}t{\rm d}\theta$ appears in the transformed metric. Although this term
looks like a rotation term in familiar spacetimes like the Kerr black hole, but in this case,
it is fictitious and lacks a physical basis.

\begin{center}
\begin{figure}[H] \hspace{4.cm}\includegraphics[width=8.cm]{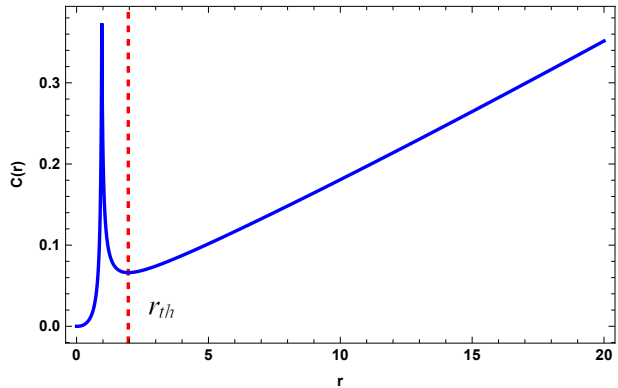}\caption{\label{cc} \small
Plot shows $C(r)$. The radius of the throat is obtained after setting $a=0.6$, $c=0.6$, $d=13$, and  $\alpha=1.2$, which corresponds to  the value $r_{th}=r_{min}=1.96$.}
\end{figure}
\end{center}

In order to see whether the new solution is still a wormhole or not, we calculate the circumference of a circle with $r, t, z=$ const, that is 
\begin{equation}
C(r)=\int ^{2\pi}_0{}g_{\theta \theta}{\rm  d}\theta=2\pi \frac{(ar)^{10/3} \cosh ^2\alpha}{r^{4/3} (a r)^{2/3} \cosh ^2\alpha  |(c+d \ln r)|+c \sinh ^2\alpha }.
\end{equation}
In Fig. \ref{cc}, we present the behavior of $C(r)$ for representative values of $a$, $c$, $d$, and $\alpha$. We can see that $C(r)$ has a minimum, which is identified as the  radius of the wormhole\rq{}s throat.

In the procedure of Kaluza-Klein reduction, the gauge fields $A_{\mu}$, and the scalar field $\psi$  can also be obtained  by using (\ref{eq8}), which are given by
\begin{equation}
A_t=-\frac{\sinh 2 \alpha \left(r^{4/3} (a r)^{2/3} |(c+d \ln r)|+c\right)}{2 \kappa 
   \left(r^{4/3} (a r)^{2/3} \cosh ^2\alpha  |(c+d \ln r)|+c \sinh ^2\alpha\right)},
\end{equation}
\begin{equation}
A_{\theta}=-\frac{a^2 r^2 \cosh \alpha }{\kappa  \left(r^{4/3} (a r)^{2/3} \cosh ^2\alpha|(c+d\ln r)|+c \sinh ^2\alpha \right)},
\end{equation}
and
\begin{equation}
\psi=-\left(\frac{c }{(a r)^{2/3}}\sinh ^2\alpha+r^{4/3} \cosh ^2\alpha| (c+d \ln r)|\right),
\end{equation}
respectively.

Substituting the gauge fields $A_{\mu}$ into the equation   $F_{\mu \nu}=\partial_{\mu}A_{\nu}-\partial_{\nu}A_{\mu}$, it will give rise to the following electromagnetic field tensors 
\begin{equation}
F_{rt}=\frac{c \sqrt[3]{r} (a r)^{2/3} \sinh 2 \alpha  (2|( c+ d \ln r)|+d)}{2 \kappa  \left(r^{4/3}
   (a r)^{2/3} \cosh ^2\alpha  |(c+d \ln r)|+c \sinh ^2\alpha \right)^2},
\end{equation}
and
\begin{equation}
F_{r\theta}=\frac{a^2 r \cosh \alpha  \left(a^{2/3} d r^2 \cosh ^2\alpha -2 c \sinh ^2\alpha
 \right)}{\kappa  \left(r^{4/3} (a r)^{2/3} \cosh ^2\alpha  |(c+d \ln r)|+c \sinh
   ^2\alpha \right)^2}.
\end{equation}

\begin{center}
\begin{figure}[H] \hspace{4.cm}\includegraphics[width=8.cm]{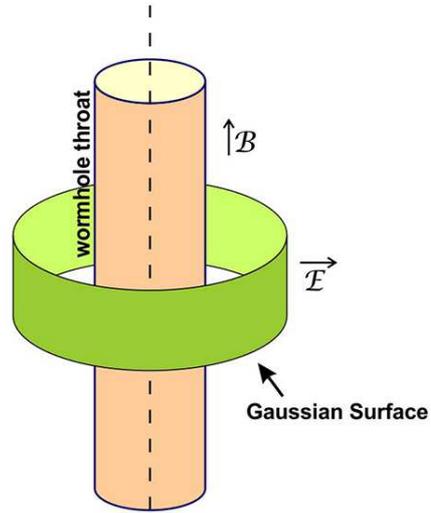}\caption{\label{wormhole} \small
Schematic diagram  of the wormhole  after boosting  with a radial electric field and a magnetic field along the $z$-axis.}
\end{figure}
\end{center}

\begin{center}
\begin{figure}[H] \hspace{4.cm}\includegraphics[width=8.cm]{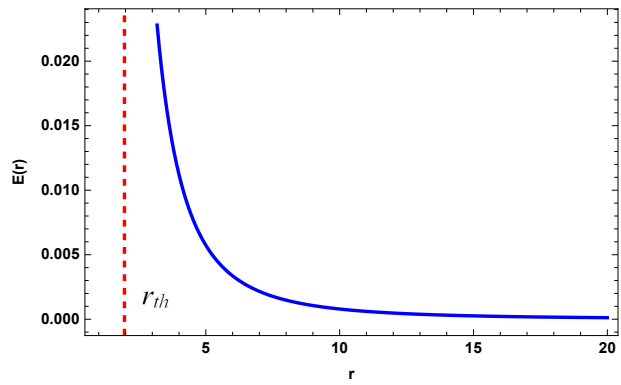}\caption{\label{E} \small
The radial electric field $E_{r}$ as a function of $r$. In the special case  $\alpha=0$, the electric field  vanishes, and we get back to the un-boosted solution.}
\end{figure}
\end{center}

\begin{figure}[H]
\centering%
\subfigure[~]
{\includegraphics[width=.41\textwidth]{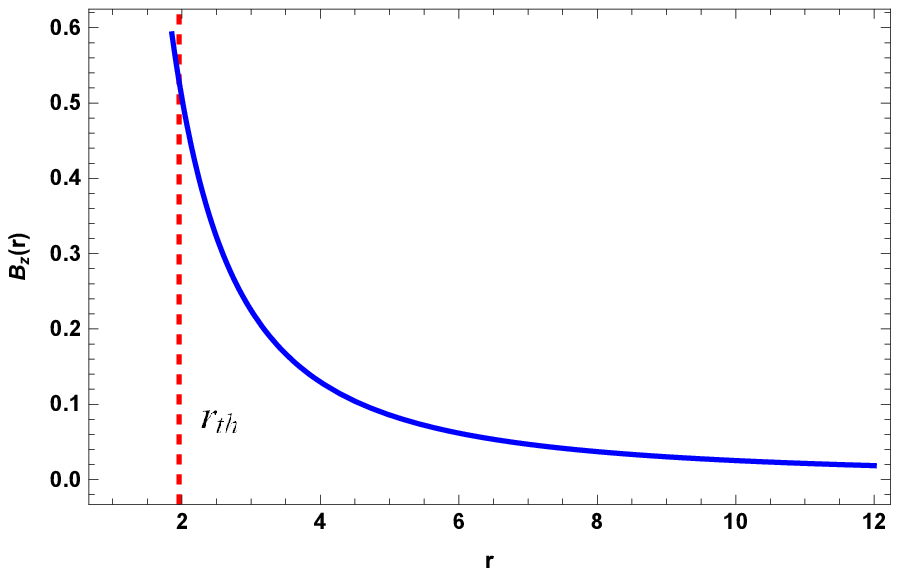} }
\subfigure[~]
{\label{fig1c}\includegraphics[width=.41\textwidth]{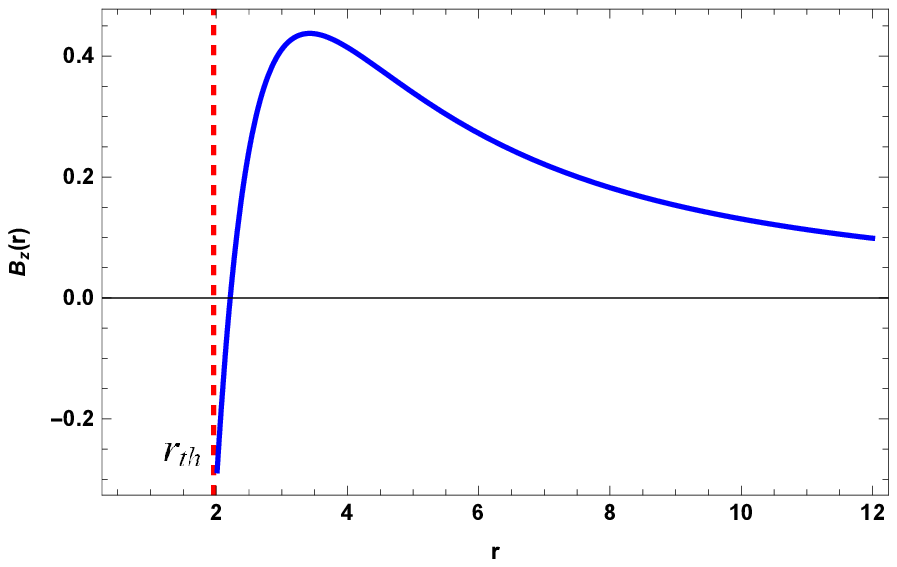} }
\caption{The plot represents the magnetic field $B_{z}(r)$ in two different cases.
\textbf{(a)}:  $a=1.4$, $c=2.4$,  $d=7$, and $\alpha=2.8$. \textbf{(b)}:  $a=0.2$, $c=1.85$,  $d=2.24$, and $\alpha=1.82$. }
\label{1}
\end{figure}

\begin{figure}[H]
\centering%
{\label{fig2c}\includegraphics[width=.41\textwidth]{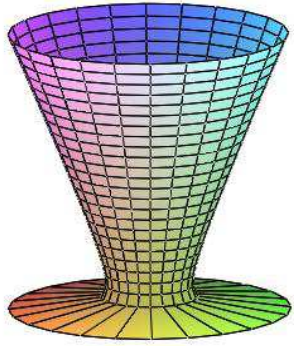} }
\caption{Visualization of the magnetic and electric wormhole. This plot is obtained by the requirement that the circumference of horizontal circles equals $C(r)$ and the vertical axis is $Z\equiv r$ with  $r\geq r_{th}$}
\label{2}
\end{figure}

It is worth noting that the  electromagnetic field tensor $F_{r \theta}$ corresponds to a magnetic field along the $z$-axis, and  $F_{rt}$  shows  a transerverse (radial) electric field, which appears  after  boost. Consequently, we have
   \begin{equation}
F^{r \theta}=\frac{B_z}{r},
\end{equation}
  \begin{equation}
F^{rt}=-E_r \rightarrow F_{r t}g^{tt}g^{rr}=-E_r.
\end{equation}
In Fig. \ref{wormhole},  an schematic of the wormhole after boosting, together with the directions of electric and magnetic fields  is shown. We also present electric field $E(r)$ and magnetic field $B_z(r)$ in Figs. \ref{E}, \ref{1}, which  both are a decreasing function of radial coordinate $r$. On the other hand, in Fig. \ref{2}, we present a visualization of the wormhole.

\section{Magnetic   and Electric  Flux }\label{M}
In this section, we derive the magnetic and electric flux through the wormhole. We will obtain the magnetic flux across the two dimensional hypersurface $t$, $z= constant$. The general expression for the magnetic flux is given by the following Gaussian integral \cite{sorkin}
\begin{align}
\Phi_{B}&=\int ^{r}_{r_{min}}\frac{1}{2}\tilde{F}^{\mu \nu}{\rm d}s_{\mu \nu}=\int ^{r}_{r_{min}}\frac{1}{2}F_{r\theta}|g^{(2)}|g^{tt}g^{zz}g^{rr}g^{\theta \theta}{\rm d}r{\rm d}\theta,
\end{align}
where ${\rm d}s_{\mu \nu}$ is an element of two dimensional surface area normal to the $z$ direction, and we take the magnetic flux integral from the location of the throat $(r_{min})$ to a distance $r$.
The  integral can not be calculated analytically, but according to our numerical calculations,  for some  ranges of the constants $a, c, d$, and $\alpha$, it  converges to a constant value.  

\begin{center}
\begin{figure}[H] \hspace{4.cm}\includegraphics[width=8.cm]{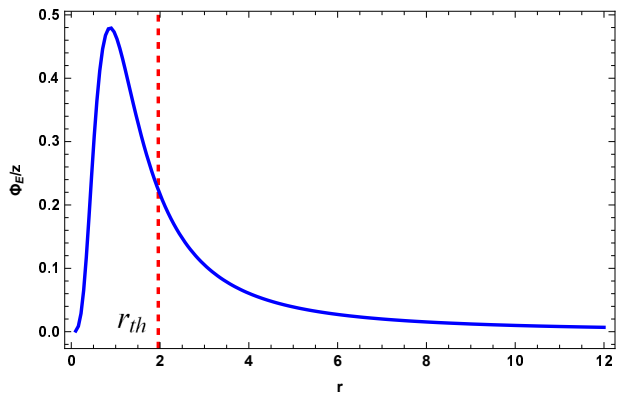}\caption{\label{f1} \small
$\Phi_{E}/z$ in the case with $a=1.28$, $c=1.8$,  $d=0.7$, and $\alpha=1.8$.  }
\end{figure}
\end{center}

Now, the electric flux per unit length (see Fig. \ref{wormhole} ) can be computed via \cite{caroll}
\begin{align}
\frac{\Phi_{E}}{z}=&\int g^{rr}g^{tt}F_{rt}\sqrt{|g^{(2)}|}{\rm d}\theta,
\end{align}
which gives the following result
\begin{align}
\frac{\Phi_{E}}{z}
=\frac{4\pi a^3 r^{10/3} \sinh 2 \alpha  \cosh ^3\alpha  (d+2 |(c+d \ln r)|)}{\left(r^{4/3} (a r)^{2/3}
   \cosh ^2\alpha  |(c+d \ln r)|+c \sinh ^2\alpha \right)^{3/2} \left(4 r^{4/3} (a r)^{2/3} \cosh
   ^2\alpha  |(c+d \ln r)|+c \sinh ^2\alpha \right)}.
\end{align}
In order to understand 
 the behavior of the electric flux, we present in Fig. \ref{f1}, the quantity  $\Phi_{E}/z$ as  a function of $r$. It can be seen from the figure that the electric charge is not localized, since otherwise we would have obtained a constant electric flux. Moreover, the electric lines of force should pass through the throat, since the electric flux does not vanish at the throat (i.e., $\Phi_{E}(r_{min}) \neq 0)$

\section{Conclusions}\label{s}
In \cite{ssh}, we presented a cylindrical five dimensional Ricci flat Kaluza-Klein solution. By using the Kaluza-Klien reduction, we derived at a four dimensional static metric, which described a wormhole. The wormhole geometry was cylindrically symmetric supported by a scalar field and a magnetic field. We further studied the magnetic flux for $r \geq r_{th}$, which converged to a constant value. We also showed that the null energy condition  was violated.

The major focus of this work is on the boosted Kaluza-Klein cylindrical and magnetized wormhole. We first apply  a boost along the extra dimension, and arrive to a     five dimensional  Kaluza-Klein   vacuum solution. By performing the Kaluza-Klein reduction, we obtain a four dimensional spacetime supported by  scalar, electric, and magnetic fields.  In order to understand the structure of the new four dimensional solution, we minimize the proper circumference of concentric circles along the Killing field $\partial_{\theta}$, and show that there is  a minimum, which  corresponds to a wormhole throat. Consequently, after the boost, and  as the most interesting outcome,  the new solution is still a  wormhole with a radial electric field, and a magnetic field oriented along the wormhole throat.

Next, we find the magnetic flux on one side of the wormhole, and show that it  converges to a constant value for some ranges of $a$, $c$, $d$, and $\alpha$. Moreover, the electric flux per unit length is  calculated, which shows that there is an extended charge distribution.\\\\

{\bf Acknowledgements}\\\\
The authors would like to thank the anonymous referee for
helpful comments. N.R. Acknowledges the support of Shahid Beheshti University.

\end{document}